\documentclass[prl,twocolumn,showpacs,preprintnumbers,amsmath,amssymb, superscriptaddress]{revtex4-1}

\usepackage[makeroom]{cancel}

\usepackage{amsmath}    
\usepackage{amssymb}
\usepackage{graphicx}   
\usepackage{verbatim}   
\usepackage{color}      
\usepackage{hyperref}   
\usepackage[normalem]{ulem}
\usepackage{natbib}

\hypersetup{colorlinks,linkcolor=blue,urlcolor=blue,citecolor=blue}
\newcommand{\beq}{\begin{equation}}
\newcommand{\eeq}{\end{equation}}
\newcommand{\bea}{\begin{eqnarray}}
\newcommand{\eea}{\end{eqnarray}}

\definecolor{darkgreen}{rgb}{0,0.5,0}
\definecolor{orange}{rgb}{1,0.5,0}
\definecolor{grey}{rgb}{.6,.6,.6}

\newcommand{\cpm}[1]{{\color{black}{#1}}}

\newcommand{\average}[1]{\langle #1\rangle}

\newcommand{\cS}{{\cal S}}

\begin{document}
\title{Quantum criticality and formation of a singular Fermi liquid in the  attractive SU($N>2$) Anderson model}
         
\author{C\u at\u alin Pa\c scu Moca}
\affiliation{BME-MTA Exotic Quantum Phases Research Group, Institute of Physics, Budapest University of Technology and Economics, 
Budafoki \'ut 8., H-1111 Budapest, Hungary}
\affiliation{Department of Physics, University of Oradea, 410087, R-Oradea, Romania}
\author{Razvan Chirla}
\affiliation{Department of Physics, University of Oradea, 410087, R-Oradea, Romania}
\affiliation{School of Medicine and Pharmacy, University of Oradea, 410087, R-Oradea, Romania}
\author{Bal\' azs D\' ora}
\affiliation{Department of Theoretical Physics and MTA-BME Lend\"ulet Topology and Correlation Research Group,
Budapest University of Technology and Economics, 1521 Budapest, Hungary}
\author{Gergely Zar\'and}
\affiliation{BME-MTA Exotic Quantum Phases Research Group, Institute of Physics, Budapest University of Technology and Economics, 
Budafoki \'ut 8., H-1111 Budapest, Hungary}

\date{\today}
\begin{abstract}
While much is known about repulsive quantum impurity models, significantly less attention has been  devoted to their attractive counterparts.
This motivated us to study the  attractive SU($N$) Anderson impurity model. While for the repulsive case, the phase diagram features mild $N$ 
dependence and the ground state is always a Fermi liquid, in the attractive case a Kosterlitz-Thouless charge localization phase transition 
is revealed for  $N>2$. Beyond a critical value of attractive interaction  an abrupt jump appears  in the number of particles at the impurity 
site, and a singular Fermi liquid state emerges, where the scattering of quasiparticles is found to exhibit  power law behavior with fractional power. 
The capacity diverges exponentially at the quantum critical point, signaling the Kosterlitz-Thouless transition. 
\end{abstract}

\maketitle

\paragraph{Introduction --} 


Quantum impurity models (QIMs)
such as  Kondo~\cite{Kondo.64} or   Anderson models~\cite{Anderson.61}
play the fundamental role of gold standards in the field of correlated systems.
In spite of their deceiving simplicity,  these models capture the nature of interactions between localized degrees of freedom and a continuum of extended states, and display a plethora of appealing effects, including
Kosterlitz-Thouless transitions~\cite{Kosterlitz_2016}, Fermi liquid vs. non-Fermi liquid behavior~\cite{Cox.98}, asymptotic freedom and quantum criticality~\cite{Vojta.2007}, just to list a few.  
They emerge in all kinds of correlated systems including  Majorana systems coupled to electrodes~\cite{Beri.2012,  Mourik.2012, Egger.2014}, Josephson junction arrays~\cite{Florens.2019}, 
or in the context of rotating molecules in a quantum liquid~\cite{Lemeshko.2017}, and they also provide the first step towards understanding bulk correlated materials~\cite{Kotliar.2006}.

In recent years, it became  possible to engineer various quantum impurity models using the artillery of nanotechnology, and to investigate correlated states and quantum phase transitions in a controlled way in these model systems
~\cite{Mebrahtu2012,Keller2013,Pierre2015,Keller2015}.
Ultracold atoms provide  further media to develop  QIMs with a tunable interaction strength~\cite{Hofstetter.06}. 
Loading ultracold atoms with large hyperfine spins into optical lattices, one can design  exotic SU($N$) spins with $N\ge 2$,  
predicted to give rise to  exotic   SU($N$)  Kondo  states~\cite{Duan.04, Paredes.05, Inaba.09, Nishida.13,Kanasz2018}. 

Most research so far focused on repulsive impurity models. However, in ultracold settings,  not only the strength but also the sign of the interspecies interaction can be tuned by using  Feshbach resonances to reach the attractive $U<0$ regime~\cite{Nishida.13}.
 Attractive interactions may also emerge in nanostructures:
 an attractive electron-electron interaction mediated by Coulomb repulsion has been demonstrated in carbon nanotubes~\cite{Hamo.16}, but attractive interactions  can also be engineered in  superconducting single electron  transistors~\cite{Averin2003}. Recent 
 experiment in heterostructures reveals the presence of 2, 3 and even 4 particle bound states~\cite{Levy.18}.
 
 In this work, we shall focus on the poorly studied $U<0$ regime of the SU($N$) Anderson model, and show that it conceals an unexpected   quantum phase transition, where charge degrees of freedom become localized.  
In this novel charge localized  phase, electrons display  singular Fermi liquid properties with power law anomalies~\cite{Zarand.2005}.
 The phase transition we find  is the impurity analogue of the  "baryonic"  transition found in attractive SU($N$) Hubbard models,  discussed intensively in the context of cold atoms~\cite{Rapp2007,Huckans.2009}.  

The SU($N$) Anderson model, we study here, describes local fermions of $N$ different flavors,  $d_{\alpha}$ ($\alpha=1,\dots,N$), 
interacting with each other  on a level of energy $\varepsilon$, and immersed in a sea of conduction electrons. 
Its Hamiltonian is defined as
\begin{equation}
\label{eq:anderson}
H  =  \frac{U}2  :Q:^2  + \varepsilon :\hat Q:
+   t \sum_{\alpha=1}^N \left(\psi^\dagger_\alpha d_\alpha+h.c. \right) +  H_{\rm bath}\;,
\end{equation}
with  $:\hat Q: = \sum_\alpha (d^\dagger_\alpha d_\alpha-1/2) = \hat Q - N/2$ the normal ordered occupation number of  level $\varepsilon$, 
and $U$ the strength of the interaction. The  term $\sim t$ describes the hybridization of the level with 
$N$ channels of conduction electrons, $\psi_\alpha$,  with the last term $H_{\rm bath}$ generating the dynamics of the Fermion field at the origin, $\psi_\alpha = \psi_\alpha(0)$. 

The standard SU(2) version of this model has  extensively been investigated  in all possible regimes~\cite{hewson} and its phase diagram is well known by now. The characteristic  behaviour of the model depends  on the position of the level $\varepsilon$, and the interaction  strength $U$ compared to the width of the level, $\Delta = \pi t^2 \varrho_0$, generated by quantum fluctuations to the electron bath, characterised by  the density of states of the  electrons at the Fermi energy, $\varrho_0 =({2\pi v_F})^{-1}$, with $v_F$ the Fermi velocity.
For large repulsive interactions, it displays the famous  spin Kondo effect~\cite{hewson,Goldhaber-Gordon1998, Kouwenhoven.2007} 
for  $-U/2 \lesssim \varepsilon \lesssim U/2$, and gives rise to mixed valence physics at $\varepsilon \approx \pm U/2$.
For $U<0$, on the other hand,  a \emph{charge} Kondo effect appears at around  $\varepsilon \approx 0$, where charge fluctuations between the states $\hat Q =0$ and $\hat Q = 2$ lead to the emergence of a Kondo resonance~\cite{Taraphder.91, Alexandrov.03, Cornaglia.04, Arrachea.05, Koch.06, Koch.07, Garate.11, Andergassen.11,Costi2012}.  
Apart from these interesting features, however, the SU(2) Anderson model displays \emph{regular} behavior for any finite  $U$ and $\varepsilon$, and assumes  a local \emph{Fermi liquid} state
with well-defined quasiparticles at low temperatures~\cite{Nozieres74, Oguri2001, Mora2015}.

\begin{figure}[t!]
	\includegraphics[width=1\columnwidth]{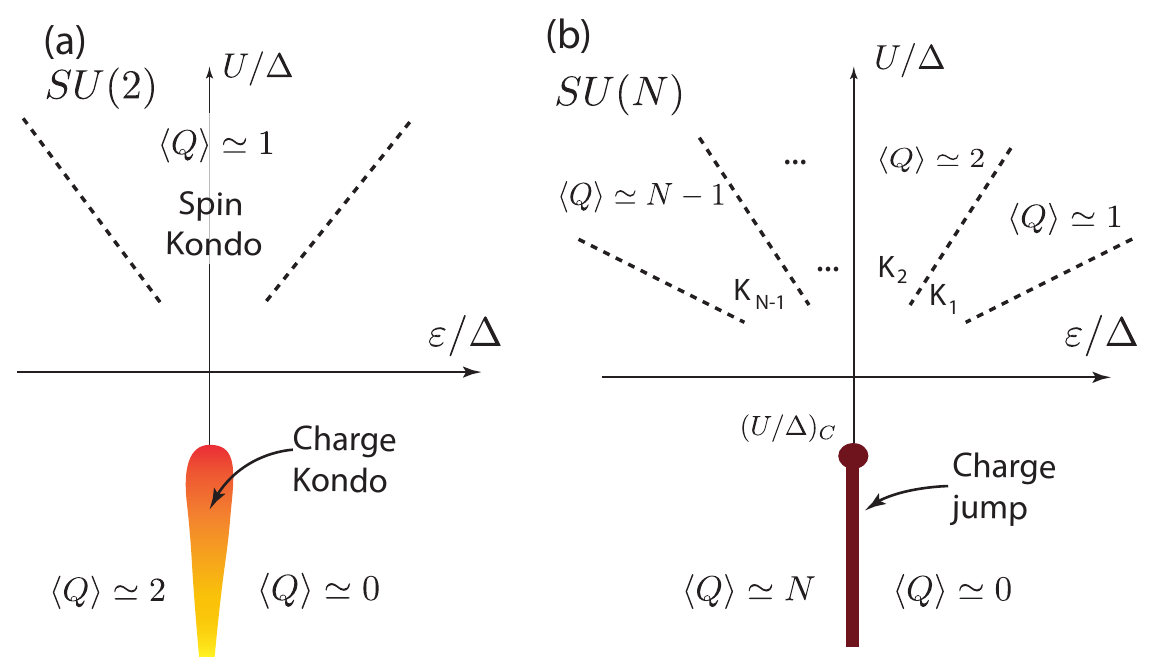}
	\caption{ (a)  Phase diagram of the SU(2) Anderson model. For positive and large 
		  $U/\Delta$  a regular SU(2) spin Kondo state (K) forms   around half-filling, $\average{Q}\simeq 1$, while at large  negative  $U/\Delta$ a charge Kondo effect emerges around $\varepsilon\approx 0$, separating the regions $\average{Q}\simeq 0$ and $\average{Q}\simeq 2$. 
		(b) The repulsive  SU(${\rm N} > 2$) Anderson model displays  a sequence of Kondo regions ($K_1,\dots,K_{N-1}$) between charging transitions  (dashed lines). For $U<0$  a charge jump
		transition appears at $U<U_C$ where  $\average{Q}$ is discontinuous  at $\varepsilon=0$. 
		 		}
	\label{fig:phase_diagram}
\end{figure}

The SU($N$) variant of the Anderson model has also attracted  a lot of attention recently due to its relevance  in various  
nanostructures~\cite{Keller2013,Hong2018,Zhang1467} and in  ultracold settings~\cite{Bloch2014}. 
Its behavior has been extensively  analyzed in the repulsive
 regime~\cite{Gorshkov.10, Moca.12, Anders.2008,Minamitani.2012, Lopez.13, Mora2008}, where it displays properties  analogous to the SU(2) case: it exhibits mixed valence fluctuations at  energies $\varepsilon \approx U(k-\frac{N-1}2)$ 
with $k=0,\dots, N-1$, and  hosts various types of SU($N$) Kondo effect between these (see Fig.~\ref{fig:phase_diagram}).  
For any $\varepsilon$ and $U>0$ one finds,  however,   a generalized  Fermi liquid state, similar to the one appearing in the SU(2) Anderson model~\cite{Mora2015}.

A naive expectation would thus be that  the attractive SU($N$) Anderson model is also a simple Fermi liquid. 
This expectation is, however, completely wrong, and, as demonstrated here,  the  $N>2$ Anderson model displays a dissipative \emph{quantum phase transition} at some critical interaction value $(U/\Delta)_c<0$,  and a singular behavior  
for $U/\Delta <(U/\Delta)_c$ (see Fig.~\ref{fig:phase_diagram}). 

\begin{figure}[b!]
	\includegraphics[width=1.0\columnwidth]{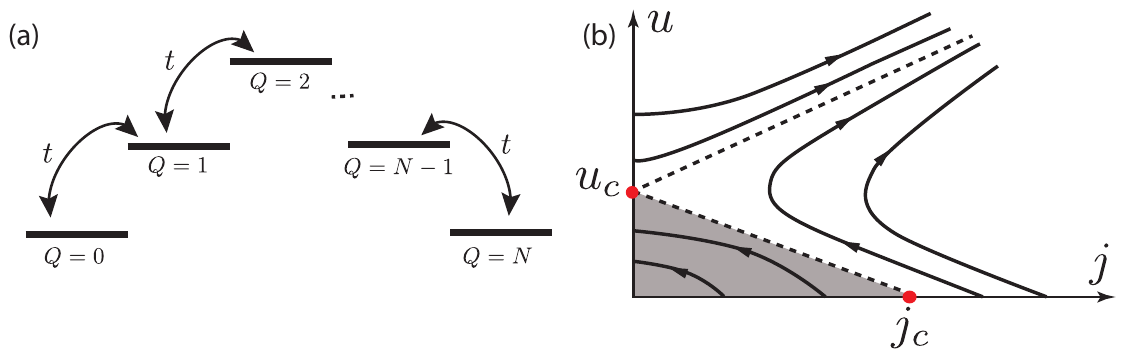}
	\caption{(a) Schematic picture of the $Q=0$ and $N$ states, connected through   high order virtual processes through $0<Q<N$ states.
(b) Leading order scaling trajectories obtained from  Eqs.~\eqref{eq:scaling}
		signaling  a Kosterlitz-Thouless phase transition at $u=u_C$ and $j=0$.   Arrows
		indicate the upon decreasing energy scale.   The shaded region marks charge localized states.}
	\label{fig:rg_flow}
\end{figure}

Before engaging in  a detailed analysis, let us give some strong and robust arguments for the existence of this phase transition. We first observe that, similar to the repulsive 
case~\cite{Yosida.70},  for $|U| \ll \Delta$ perturbation theory in $U$ is governed by the expansion parameter  $|U| /\Delta \ll 1$, and therefore  all properties are analytical functions of $U$ and $\varepsilon$. 
Thus the Fermi liquid state at $U>0$ naturally extends to the regime of small negative $U$'s.
The situation is, however, dramatically different for $|U|\gg \Delta$. Focusing  on the regime, $\varepsilon\approx 0$, we may want to attempt there to perform perturbation theory in $t$. For $\varepsilon =0$ and $t=0$, however, the lowest lying $\hat Q =0$ and 
$\hat Q =N$ states of the isolated level are  degenerate. As depicted in Fig.~\ref{fig:rg_flow}.a, 
these two states are connected through   high order virtual processes yielding a "tunneling"  term $\sim (
D^\dagger \psi_1\dots \psi_N + \text{h.c.})$ with $D^\dagger$ defined as $D^\dagger \equiv d_N^\dagger \dots d^\dagger_1$
(see Fig.~\ref{fig:rg_flow} (a)). 
Simple power counting  shows that this tunneling term is \emph{irrelevant} in the renormalization group sense. As a consequence, charge fluctuations 
between the states  $\hat Q =0$ and $\hat Q =N$ must be suppressed at zero temperature. This leads us to the conclusion that for large attractive interactions, a first order "charge jump"  quantum phase transition must take 
place, as depicted in Fig.~\ref{fig:phase_diagram}.

To substantiate this claim and describe this phase transition in detail, we  first identify SU($N$) invariant terms generated by quantum fluctuations of the charge, $\hat Q$. 
In addition to the tunneling term discussed above, charge fluctuations $\hat Q = 0 \leftrightarrow 1$ and $\hat Q = N \leftrightarrow N-1$  generate a charge state dependent potential
 of the form   $\sim (D^\dagger D-1/2)\sum_\alpha \psi^\dagger_\alpha\psi_\alpha$. This leads us to   the effective Hamiltonian  
\begin{eqnarray}
H_{\rm eff} &=& N  \varepsilon \;(D^\dagger D -{1\over 2})  +  j\; \cpm{v_F} \,a^{N/2-1}
(\psi^\dagger_1\dots \psi^\dagger_N D +\text{h.c.}) 
\nonumber
\\
& + & u\,\big(D^\dagger D-{1\over 2}\big)\sum_{\alpha=1}^{N}  v_F\, \psi^\dagger_\alpha \psi_\alpha  
+H_{\rm bath} \label{eq:H_eff}
\end{eqnarray}
with $j\sim t^N/|U|^{N-1}$ denoting  a dimensionless amplitude    of  "charge exchange",  $a\sim v_F/U$ the natural cut-off parameter of the problem, and $u\approx 4\Delta/(|U|(N-1))$ the dimensionless strength of potential scattering.  
The fermionic bath   in Eq.~\eqref{eq:H_eff} can be expressed  in terms of chiral one-dimensional fermions, 
$H_{\rm bath} = v_F \sum_{\alpha =1}^N \int \psi_\alpha^\dagger(x) \,i \partial_x \psi_\alpha(x) \,{\rm d} x/(2\pi)$, 
with $\psi (x> 0)$ and $\psi (x< 0)$ representing incoming and outgoing spherical waves~\cite{Affleck1993}, 
respectively~\footnote{Notice the slightly unusual real space normalization of the field $\psi$.}.

A straightforward renormalization group analysis of $H_{\rm eff}$ then yields the scaling equations 
(see the Supplemental material)
\begin{eqnarray}
\frac{{\rm d} j}{{\rm d} \ln(a)} & = & \Big(1-{N \over 2}\Big) j +  N\,j \,u + \dots\;, \nonumber\\
\frac{{\rm d} u}{{\rm d} \ln(a)}& = &  4\;j^2+\dots \;,
\label{eq:scaling}
\end{eqnarray} 
and the flow diagram presented in Fig.~\ref{fig:rg_flow} (b).  Clearly, below  a critical value, 
$u<u_C$,  (i.e., $\Delta/|U|<(\Delta/|U|)_C$) a small charge exchange $j$ is irrelevant, while for $u>u_C $, i.e., $\Delta/|U|>(\Delta/|U|)_C$, one always finds a
flow to strong coupling, $j\to\infty$, signaling the dominance of  quantum fluctuations and the formation of a charge Kondo state. The phase transition is of Kosterlitz-Thouless type,  implying a  Fermi liquid scale vanishing   as
\begin{equation}
T_{\rm FL}\sim \exp\Big(- C \sqrt{  \frac{|U|}{\Delta-\Delta_C}}\Big)\;.
\label{eq:Tstar}
\end{equation}

\paragraph{Mapping to the dissipative two-state system. --} 
The nature of the  phase transition  can  further be clarified  by means of Abelian bosonization, which we use to map our model 
\eqref{eq:H_eff} to an Ohmic dissipative two state system~\cite{Legget1987}. To this end, we express  
the fermionic fields  in terms of chiral bosons,  $\psi_\alpha(x) \equiv (\gamma_\alpha/ \sqrt{a})\, e^{-i\varphi_\alpha(x))}$, with the Klein factors $\gamma_\alpha$ -- now represented as Majorana fermions -- assuring 
the correct anticommutation relations between different Fermion fields. Introducing now the charge field, $\varphi_c \equiv (\varphi_1 + ... +\varphi_N)/\sqrt{N}$, we   express the Hamiltonian as 
\begin{eqnarray}
H_{\rm eff}   &=& H_{\rm bath}  
	+ j  \frac {v_F} {a} \big( \gamma  \,D \, e^{i \sqrt{N}\varphi_c(0)} + \text{h.c.}\big) \nonumber
	\\
	&+& u \; v_F \; \sqrt{N}\,(D^\dagger D -{1\over 2})\, \partial_x\varphi_c(0) \;,
	\label{heffboson}
\end{eqnarray}
with $\gamma \equiv \gamma_1 ... \gamma_N$. 
The fermionic bath can be represented here as 
 $$
 H_{\rm bath} = {v_F\over 4\pi}\int :(\partial_x\varphi_c(x))^2: dx + ...\;,
 \nonumber \\
$$
with the dots referring to  SU($N$) spin excitations, decoupled from the local charge degree of freedom. 
In the subspace of  $D^\dagger |0\rangle$ and $ |0\rangle$, the operators $D^\dagger \gamma$  and $D^\dagger D -{1\over 2}$ act as  the usual Pauli operators, $\sigma^+$ and $\sigma^z/2$. A further 
unitary transformation with $U= e^{i\sqrt{N}\varphi_c(0) \sigma_z /2}$ eliminates in $H_{\rm eff} $ the 
phase factors while shifting $u\to \tilde u = u - 1$, and yielding Leggett's Ohmic dissipative 
two-state system model,  
\begin{eqnarray}
\tilde H_{\rm eff}   &=& H_{\rm bath}  + \frac{\Delta}2 \sigma_x  -  \sqrt{\frac{\alpha} 2}v_F\;  \sigma_z \; \partial_x \varphi_c\;.
\end{eqnarray}
Here $\Delta = 2j v_F/a$ denotes the  tunneling amplitude of the two-state system, 
while  $\alpha = {N\over 2}(1 - u)^2$ stands for  Leggett's famous dissipation 
parameter~\footnote{In the bosonization scheme, $u$ is identified as $u=2\delta/\pi$, with $\delta$ the scattering 
phase shift experienced by the fermions $\psi_\alpha$.}.
The latter parameter governs the dissipative phase transition 
between a charge delocalized phase for $\alpha<1$  with coherent charge oscillations for $\alpha<1/2$,  and a charge localized state for $\alpha>1$.  We thus conclude that the charge localization transition in the $U<0$ Anderson model 
is essentially identical to Leggett's famous \emph{dissipative} quantum phase \emph{transition}, with  the critical value, $\alpha=1$
corresponding to $u_c = 1-\sqrt{2/N}$
~\footnote{It also bears certain similarity to dissipation induced charge localization 
transitions in heavy fermion compounds~\cite{Watanabe2006,Borda2006}}.


\paragraph{NRG approach.---} 
Our discussion above follows from  an effective model, strictly justified only in the limit   $|U|\gg \Delta$, a condition 
 certainly violated  in the vicinity of the predicted quantum  phase transition. To substantiate our claim and to demonstrate the Kosterlitz-Thouless character of the  phase transition, 
we  resort to (density matrix) numerical renormalization group (NRG), which we apply directly on  the Anderson model \eqref{eq:anderson} in the first non-trivial  cases, $N=3$ and $N=4$. Then the Hamiltonian~\eqref{eq:anderson}  has  a global U(1)$\times$ SU($N$) symmetry
corresponding to charge conservation and SU(3) spin rotations,  which we both exploit in our computations~\cite{BudapestNRG, Moca.12}.

%
%

\begin{figure}[t!]
\includegraphics[width=0.8\columnwidth]{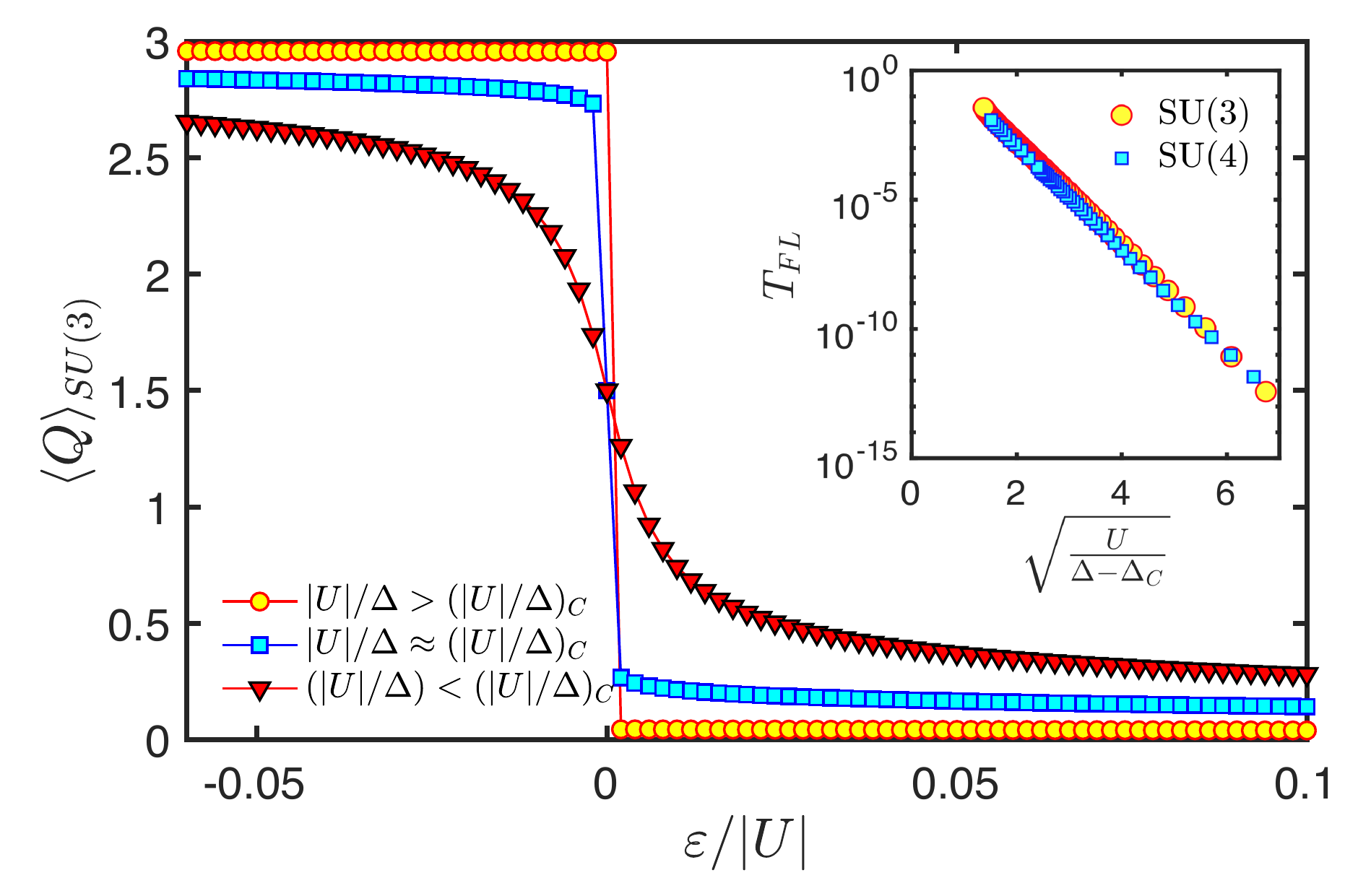}
\caption{  The occupation number $\langle \hat Q\rangle$ as function of $\varepsilon /|U|$ for 
 different ratios of $|U|/\Delta$. 
Inset: Evolution of the Fermi liquid scale $T_{\rm FL}$ 
along the charge jump line for $|U|/\Delta< (|U|/\Delta)_C$.
}
\label{fig:occupation}
\end{figure}

The average occupation $\average{\hat Q} =\sum_\alpha \langle d^\dagger_\alpha d_\alpha  \rangle$
is presented in Fig.~\ref{fig:occupation} for $N=3$ as a function of $\varepsilon$  for a fixed $U<0$ but different values of $\Delta$. 
As expected, for $\varepsilon\to\infty $,  the local level becomes  empty, while for $\varepsilon \to-\infty$, it becomes completely occupied. 
For small attractive interactions, the occupation $\average{\hat Q}$ gently crosses over between these two values as 
  $\varepsilon$ is varied, and is exactly $\langle \hat Q\rangle =3/2$ at  the particle-hole  symmetric point, $\varepsilon = 0$.
The Kosterlitz-Thouless phase transition is signaled by the appearance of a sudden jump in $\average{\hat Q}$ once the ratio $U/\Delta $ exceeds  a  critical value, $(|U|/\Delta)_c\approx{ {2.43}}$~\cite{footnote1}.

The   capacity, $\chi=\partial_\varepsilon \langle \hat Q\rangle_{\varepsilon =0}$ diverges as one approaches the transition from  the Fermi liquid side, $|U|/\Delta < (|U|/\Delta)_C$, and its inverse defines the Fermi liquid scale, $T_{\rm FL}\equiv \chi^{-1}$.
The  scale   $T_{\rm FL} $ is found to  vanish exponentially as one approaches the transition from the weak coupling side, 
in agreement with  Eq.~\eqref{eq:Tstar}  (see the  inset of Fig.~\ref{fig:occupation}).

\paragraph{Scattering and singular Fermi liquid.---} 

The Fermi liquid scale vanishes on the first order transition line, $\varepsilon=0$  and   $|U|/\Delta >(|U|/\Delta)_C$, 
where a free charge degree of freedom appears. It is this residual charge degree of freedom, which is responsible for the 
singular Fermi liquid properties.

\begin{figure}[t!]
\includegraphics[trim=0cm 0cm 1.5cm 1cm, clip=true,  width=1\columnwidth,]{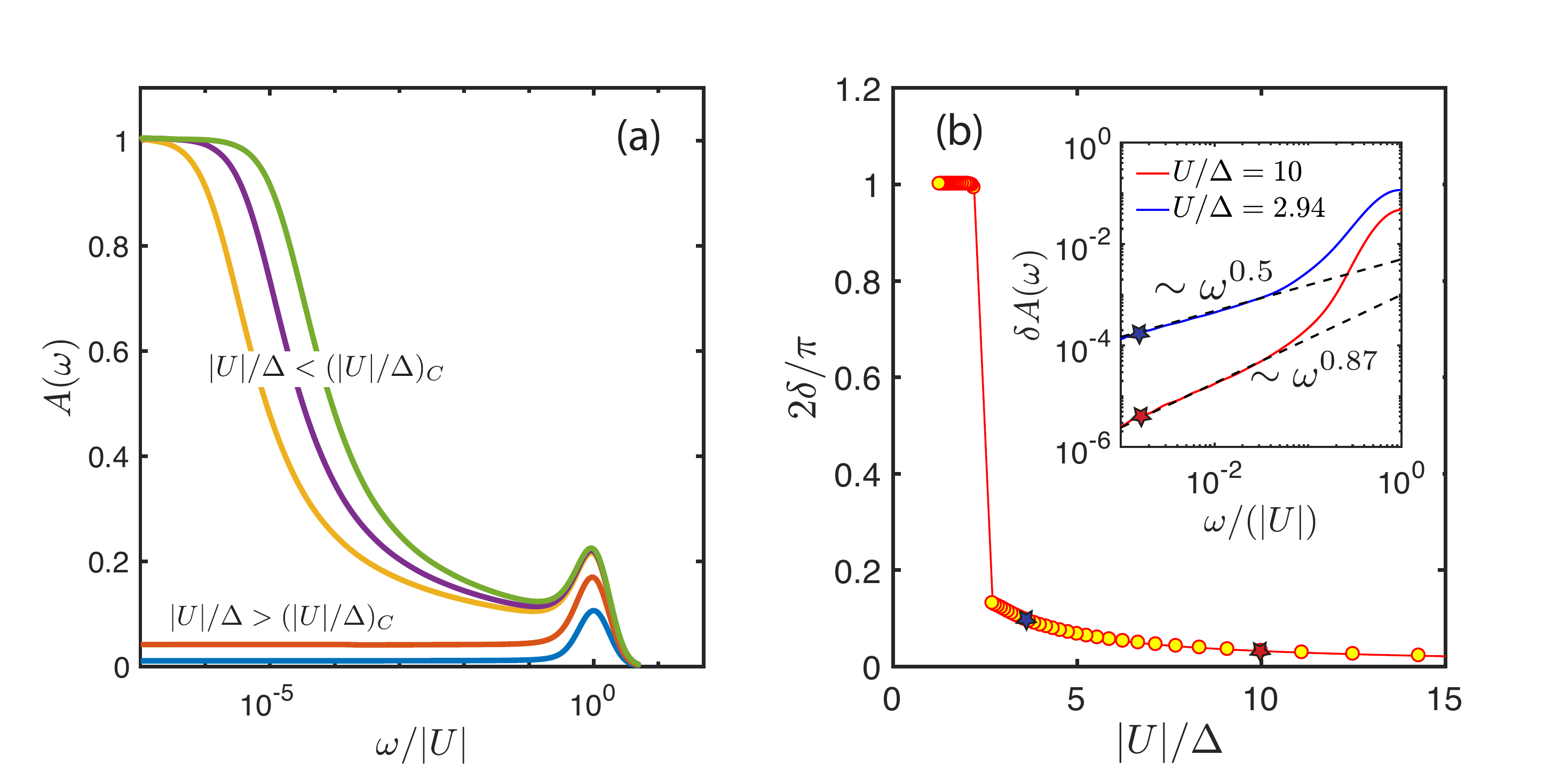}
  \caption{(Color online) (a) Evolution of the 
  dimensionless scattering cross section  $A(\omega)$ for $N=3$ as $|U|/\Delta$ is increased beyond its critical value, 
   $(|U|/\Delta)_c\approx 2.43$. For  $|U|/\Delta < (|U|/\Delta)_c$ a charge Kondo resonance is observed below the  Fermi liquid 
   scale, and a corresponding maximally strong scattering, while for $|U|/\Delta > (|U|/\Delta)_c$ a singular Fermi liquid state emerges, 
   with a small residual scattering cross section,  and a power law correction. 
   (b) Evolution of phase shift across the transition. The inset displays the scaling of the spectral function for $|U|/\Delta= 2.94$
   and $|U|/\Delta = 10$.  With increasing $U$, the exponent $\beta$ approaches 1 as expected from bosonization. } 
\label{fig:A_w}
\end{figure}

The charge localization transition is also clearly observable in the scattering properties of the carriers. We
 have  determined numerically the energy dependence of  the total   scattering cross section $\sigma(\omega)$ of the conduction electrons 
as a function of their energy, $\omega$. By the optical theorem,   $\sigma(\omega)$ 
is just proportional to   the imaginary part of the T-matrix, $\sigma(\omega) \propto -{\rm Im}T(\omega)$, which is directly proportional to the $d$-level's propagator,  $ T(\omega) = 2\pi v_F  \Delta\; G_{d_\alpha d_\alpha^\dagger}(\omega)$.   Fig.~\ref{fig:A_w} displays  the dimensionless cross section  $ A(\omega) \equiv -  {\rm Im}\,T(\omega)/(2\pi v_F)$ below and above the quantum phase transition.

In the Fermi liquid phase, scattering becomes maximally strong at  energies below the Fermi liquid scale, $ T_{\rm FL}$, 
and the dimensionless cross section exhibits an analytical behavior, $A(\omega)=  1 - C\, \omega^2/T_{\rm FL}^2$+\dots. In contrast, in the charge localized phase  $A(0)$ takes on a small value, $A_0= \sin^2(\delta/\pi)$,  
 practically independent of the energy of the electrons. 
Here $\delta$ denotes the charge state dependent residual phase shifts, characterizing the scattering of electrons at the Fermi energy. These are directly related to the interaction parameter $u$ in the bosonization approach, $\delta \approx \pm u~\pi /2$. This residual phase shift, extracted from the finite size spectrum of our  NRG calculations, is displayed in Fig.~\ref{fig:A_w}. It assumes 
 a universal  value $\delta_c  = (1-\sqrt{2/N})\; \pi / 2 $   at the transition point, and  tends to   $\delta \to  0$  in the limit of  strong attractive interactions, $|U|/\Delta\to\infty$. 
Notice that the critical value, 
$A_{c}(0) \approx 0.081$ (also confirmed  by our numerics) is substantially reduced compared to the maximally strong scattering characterizing the Fermi liquid phase, implying that  charge localization reduces the scattering cross section by $\sim 92\%$.

In the charge localized phase, the scattering cross section displays an additional, slow power law dependence, $A(\omega) = A_0 + A_1 
|\omega|^\beta+\dots$.  The exponent $\beta$ here can be determined from  our effective model, \eqref{eq:H_eff}, by means of Abelian bosonization.
For $N=3$ this yields $\beta =\beta_c = 2(1 - \sqrt{2/3})$ at the transition point, with $\beta$ gradually increasing  to $\beta =1$ in the limit of 
strong attractive interaction  (for the precise relation, see the Supplemental material~\cite{SuppMat}).   
This dependence is indeed verified by our NRG computations, shown in the inset of Fig.\ref{fig:A_w}.b.

\paragraph{Conclusions.--}
In this work, we have demonstrated that one of the most paradigmatic models, the attractive SU($N$) Anderson model displays a surprising charge localization transition for $N>2$,  characterized by an exponentially divergent Fermi liquid scale, a jump in the zero-temperature scattering cross section, and an emerging singular Fermi liquid state. 
The phase transition analyzed here is thus the impurity analogue  of the baryon - color superfluid phase transition predicted 
in the attractive SU(3) Hubbard model~\cite{Rapp2007}, realized by trapped  $^6$Li atoms. 
The charge localized phase of the Anderson model corresponds simply to  the baryonic state, where heavy particles are bound together to trions, while the attractive heavy Fermi liquid  formed at 
$0> U/\Delta > -(|U|/\Delta)_C$  is the mirror image of the color superfluid. 

The present  results signal a similar transition  in the attractive SU(4)  Hubbard model, 
where quadruplons (4 particle bound states) must  form at strong attractive interactions, 
while at weak couplings, a superfluid state is the natural candidate. 
 These quadruplons may form  a strongly interacting  Bose condensate or 
display  charge localization.  We  remark that  bound  $N$-ons become very heavy, since their effective hopping rate on a lattice is renormalized as 
$t_{\rm eff}\sim t^N/|U|^{N-1}$, and is much smaller than  the  nearest neighbor repulsion $ V\sim t^2/|U|$ generated by quantum fluctuations. Therefore, attractive  strongly  interacting SU($N$) systems are susceptible to charge ordering, in close analogy with  valence skipping compounds~\cite{Dzero2005}.  

\emph{Acknowledgments.--} We thank Alexei Tsvelik and Eugene Demler for insightful  discussions.  
This work has been supported by the National Research, Development and Innovation Office (NKFIH) through Grant No.  K119442 and through  the Hungarian Quantum Technology National Excellence Program, project no. 2017-1.2.1-NKP-2017- 00001, and by  the  Romanian National Authority for Scientific Research and Innovation, UEFISCDI, under 
project no. PN-III-P4-ID-PCE-2016-0032.

\bibliography{references}

\begin{thebibliography}{64}
\bibitem [{foo()}]{footnote1}%
  \BibitemOpen
\bibfield  {journal} {  }\href@noop {} {\bibinfo  {journal} {For the SU(4) we
  find a critical value $(|U|/\Delta)_C\approx{12.18}$}\ }\BibitemShut
  {NoStop}%
\end{thebibliography}

\newpage
\null
\section{Supplemental Information}

\subsection{Effective model and the perturbative RG}

We can treat our effective model defined in Eq.~(2) of the main text 
by using  Wilson's perturbative renormalization group (RG) approach~\cite{WilsonRG}. For that
we express the partition function in the path integral language as 
\begin{equation}
Z = Z_0 \cdot \langle e^{- {\cal S}_\text{int}}\rangle_0\,,
\end{equation}
where $ {\cal S}_\text{int}$ denotes the interaction part of the action, 
\begin{widetext}
\begin{equation}
 {\cal S}_\text{int} = \int_0^\beta \text{d}\tau\,
  \Bigl[ j v_F a^{N/2-1}
\big(\overline \psi_1(\tau)\dots \overline\psi_N(\tau) D(\tau) +\text{h.c.}\big)
+  
u \,v_F\,\sum_{\alpha=1}^{N} \big(\overline D(\tau) D(\tau)-{1\over 2}\big)   \overline\psi_\alpha(\tau)  \psi_\alpha  (\tau) 
+\dots \bigr] \;,
\label{eq:Sint}
\end{equation}
\end{widetext}
$Z_0$ is the unperturbed partition function, and $\langle\dots\rangle_0$ indicates 
averaging with the non-interacting action, 
${\cal S}_0$. The interaction-induced change in the Free energy is then just given by $\delta F = -T \ln \, \langle 
e^{-{\cal S}_\text{int}}\rangle_0$, while imaginary time correlation functions can be generated by adding appropriate source terms to 
the action.

The free energy can be computed by expanding $\langle 
e^{-{\cal S}_\text{int}}\rangle_0$ systematically in 
the interaction terms, and then using Wick's theorem to  evaluate contractions in the expansion  in terms of the 
composite fermion's unperturbed propagator, ${\cal G}_D (\tau) \equiv \langle D(\tau)\overline D(0)\rangle_0= \frac 1 2 \,
\text{sgn}(\tau)$, and that of the conduction electrons,  ${\cal G} (\tau) = \langle\psi_\alpha(\tau)\overline\psi(0)\rangle_0$, 
 \begin{equation}
{\cal G} (\tau)= \frac{\pi/\beta}{v_F\, \sin ({\pi\over\beta} (\tau +(a/v_F)\,{\rm sgn}\tau))}.
\end{equation}

In Wilson's RG scheme, the renormalization group equations are  generated by requesting the invariance of all physical observables under the rescaling of $a$ and the couplings $u$ and $j$. Ultraviolet divergencies must be handled properly during this process.  From the RG point of view, $j$
is an irrelevant coupling that scales to zero at the tree level. Still, along the RG procedure, it generates 
marginal terms  $\propto u$, which renormalize  $j$ itself, and can ultimately drive $j$ to strong coupling. 

\begin{figure}[tbhp!]
	\includegraphics[width=0.95\columnwidth]{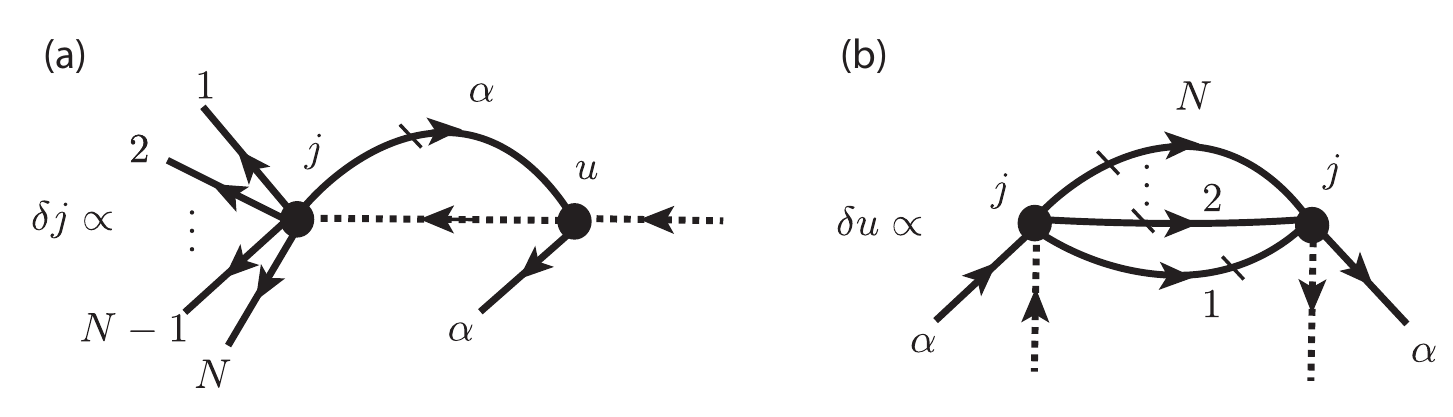}
	\caption{ Feynman diagrams for the second order corrections in the perturbative 
		RG. Crossed lines indicate 
		a logarithmic derivative with respect to the scaling parameter $a$.
		Diagram (a) renormalizes the exchange coupling $j$ while diagram (b) renormalizes $u$.
	}
	\label{fig:rg_diagrams}
\end{figure}

The leading order transformation can be read out directly from the structure of  
 \eqref{eq:Sint}, yielding the linear terms in Eq.~(3) of the main text. To  capture the quantum phase transition, it is sufficient to restrict ourselves to  second order terms, i.e. to one loop order.  The relevant second order contractions  are presented in Fig.~\eqref{fig:rg_diagrams}.  
The diagram in panel (a) in Fig.~\ref{fig:rg_diagrams} renormalizes the 
exchange coupling $j$ while 
the one in panel (b) renormalizes $u$. 
For example, by rescaling $a$,  the   diagram depicted in Fig.~\ref{fig:rg_diagrams}(a) generates a term in the renormalized  effective action 
\begin{equation}
\delta \cS_{\rm int} =  - N v_F^2 \,\cpm{a^{N/2-1}}\, (u j)\int \text{d}\tau \prod_{\alpha=1}^N\bar\psi_\alpha(\tau)\ D(\tau)
\int \text{d}\tilde \tau \delta{\cal G}(\tilde \tau)\; ,
\end{equation}
where the prefactor $N$ comes from the internal contractions, and $\delta{\cal G}$ denotes the change in the propagator induced by the rescaling of $a\to a'$. Observing that   $\delta{\cal G}$  falls off rapidly, the time arguments of 
the fields can be made equal, and the contraction can be replaced by its integral, 
  $\int dt' \delta{\cal G}_{\alpha}(\tau) \approx - \cpm{v_F}^{-1} \,2 \delta a / a $. 
Clearly, the term generated this way can be incorporated in the original action by simply  
renormalizing the couplings.  In terms of  the scaling variable,
${\rm l}=\ln(a/a_0)$, this gives us the second order terms of the differential equations 
presented in the main text,  Eqs.~(3).

\vspace{0.5cm}
\subsection{Low frequency properties of the T-matrix in the charge localized phase} 

The asymptotic behavior of the scattering cross-section can be determined by means of Abelian bosonization. 
By investigating the equation of motion of the propagator $G_\alpha(t) = -i \langle T \psi_\alpha(t)\psi_\alpha^\dagger(0)\rangle$  and the fields $\psi_\alpha$ and $\psi^\dagger_\alpha$, we find that the self-energy of 
the propagator $G_\alpha(t) = -i \langle T \psi_\alpha(t)\psi_\alpha^\dagger(0)\rangle$ is simply the  correlator
$$ 
\Sigma_\alpha(t) = -i \langle T_t O_\alpha(t)O^\dagger_\alpha(0)\rangle,
$$
where $O_\alpha$ stands for the composite  operator, $O^\dagger_\alpha =  j\; a^{N/2-1} D^\dagger \psi_N \dots \psi_{\alpha+1} \psi_{\alpha-1} \dots \psi_1$.
Within the effective field theory, Eq.~(2) in the main text, 
this operator corresponds to the  operator $d_\alpha^\dagger$.
For simplicity, let us focus on $O^\dagger_1$. In the bosonized form, this operator can be expressed as 
$j a^{-1/2}  D^\dagger\gamma_N\dots\gamma_2 e^{-i (\varphi_2+\dots + \varphi_N)}$. Next we introduce the charge field 
$\varphi_c \equiv \overline \varphi\cdot {\overline e}_c $ with ${\overline e}_c = \{1,1,\dots,1\}/\sqrt{N}$, and $N-1$ orthogonal and properly normalized spin fields,  $\varphi_s^{(k)} \equiv \overline \varphi \cdot{\overline e}^{(k=1,..,N-1)}_{\;s} $  with the ${\overline e}^{(k)}$ denoting unit vectors orthogonal to ${\overline e}_c $  and to each other.

We can then rewrite the operator $O^\dagger_1$ as
\begin{equation}
O^\dagger_1 = j \;a^{-1/2}  D^\dagger \gamma_N\dots \gamma_2 \exp\{-i (\frac {N-1}{\sqrt{N}} \varphi_c + q \;\tilde \varphi_s)\}
\label{eq:O}
\end{equation}
with $q^2 =\frac{N-1}N$ and $\tilde  \varphi_s$ a properly normalized combination of the spin fields $\varphi_s^{(k)}$. 
Clearly, only the charge part of this operator is affected by the interaction terms in Eq.~(5) in the main text. 

By the optical theorem,  the scattering cross section at energy $\omega$ is directly proportional to the 
imaginary part of the retarded self energy at that frequency. Therefore, we need to determine the low frequency 
behavior  of $\Sigma_\alpha(\omega)$, related to the long time behavior of $\Sigma_\alpha(t)$. 
Since the coupling $j$ is irrelevant along the charge transition line, in leading order, we can set it to zero in Eq.~(5), and then eliminate the term $\sim u$ by the unitary transformation  $\tilde U= e^{i u \varphi_c(0) \sigma_z /2}$, transforming $O^\dagger_1$ into 
\begin{equation}
 \tilde O^\dagger_1 \sim D^\dagger \gamma_N\dots \gamma_2 \; e^{-i \big(\frac {N-1}{\sqrt{N}}-\sqrt{N} u\big)\tilde  \varphi_c -i  q \;\tilde \varphi_s}\;.
\label{eq:tildeO}
\end{equation}
with $u$ related to the residual phase shift as $u=2\delta/\pi$.  Since, at this point,  we have eliminated all non-trivial 
terms of the Hamiltonian, the asymptotic behavior is just determined by the vertex operator in \eqref{eq:tildeO}, 
yielding 
$$
\Sigma_1(t)\sim \ {t^{-(N-1-2u(N-1) + u^2 N)}}\;\;. 
$$
Fourier transformation then leads straightforwardly to the expression,
$$
\beta = N-2-2u(N-1) + u^2 N\;.
$$

\end{document}